\documentclass[%
 reprint,
 twocolumn,
showpacs,
 amsmath,amssymb,
 aps,
 prl,
]{revtex4-1}

\usepackage[usenames,dvipsnames,svgnames,table]{xcolor}
\usepackage{graphicx}
\usepackage[colorlinks=true,
            linkcolor=blue,
            urlcolor=blue,
            citecolor=blue,
            urlbordercolor = white,
            ]{hyperref}
\usepackage[utf8]{inputenc}
\usepackage[english]{babel}

\addto\captionsenglish{}

\usepackage{lipsum}
\def\p{{\partial}}
\def\Re{\mathop{\text{Re}}}
\def\Im{\mathop{\text{Im}}}

\RequirePackage{color}

\begin{document}

\title{Quantized Laplacian growth, III: On conformal field theories of Laplacian growth}
\author{Oleg Alekseev}
\email{teknoanarchy@gmail.com}

\affiliation{%
Chebyshev Laboratory, Department of Mathematics and Mechanics, Saint-Petersburg State University, 14th Line, 29b, 199178, Saint-Petersburg, Russia.
}%

\date{\today}

\pacs{05.10.Gg, 68.05.-n, 47.52.+j, 11.25.Hf}
\begin{abstract}
A one-parametric stochastic dynamics of the interface in the quantized Laplacian growth with zero surface tension is introduced. The quantization procedure regularizes the growth by preventing the formation of cusps at the interface, and makes the interface dynamics chaotic. In a long time asymptotic, by coupling a conformal field theory to the stochastic growth process we introduce a set of observables (the martingales), whose expectation values are constant in time. The martingales are connected to degenerate representations of the Virasoro algebra, and can be written in terms of conformal correlation functions. A direct link between Laplacian growth and the conformal Liouville field theory with the central charge $c\geq25$ is proposed.
\end{abstract}

\maketitle

Nonlinear growth phenomena still pose great challenges in nonequilibrium statistical physics and mathematics. The growth processes observed in various physical, chemical and biological systems typically lead to the formation of self-similar patterns with remarkable geometrical properties~\cite{StanleyBook,PelceBook}. Most of them fall naturally into universality classes depending on the mechanism driven the growth. However, even the problems from the same class often require absolutely different approaches to study them.

The relevant examples are the \textit{deterministic} dynamics of the interface in a Hele-Shaw cell~\cite{BensimonRMP}, and discrete \textit{stochastic} fractal growths, such as diffusion-limited aggregation~\cite{WittenSanderPRL}. Although these phenomena are both diffusion driven growth processes, the stochastic dynamics is mainly studied numerically, while the deterministic Hele-Shaw problem is known to possess a reach mathematical structure~\cite{Richardson,WiegmannPRL}. A single framework unifying these processes must resolve the following obstacles: i) A naive limit of the vanishing particle size in diffusion-limited aggregation leads to the ill-defined Hele-Shaw problem, when the interface quickly develops cusps, and ii) There are no obvious ways to incorporate noise in the deterministic Hele-Shaw dynamics.

Both obstacles can be resolved by introducing a short-distance cutoff~$\hbar$, so that the change of areas of domains is quantized and equals an integer multiple of the area quanta $\hbar$~\cite{QLG1}. The cutoff prevents the cusps production, and generates inevitable noise on a microscale. The mean field approach allows one to reduce noise to a set of collective coordinates---specific singularities of the uniformization map of the domain~\cite{QLG2}. However, this approach generally fails to take account of quantum fluctuations of the collective motion~\cite{Lacroix14}.

In this paper, we use the \textit{stochastic mean field method} to study the interface dynamics in quantized Laplacian growth by adding noise to the motion of collective coordinates. The backbone of the theory is an assumption that the quantum dynamical problem, that is the evolution of the quantized domain in the Laplacian growth process~\cite{WiegmannQH,QLG1}, can be replaced by a superposition of classical dynamics.

\begin{figure}[t]
\centering
\includegraphics[width=1\columnwidth]{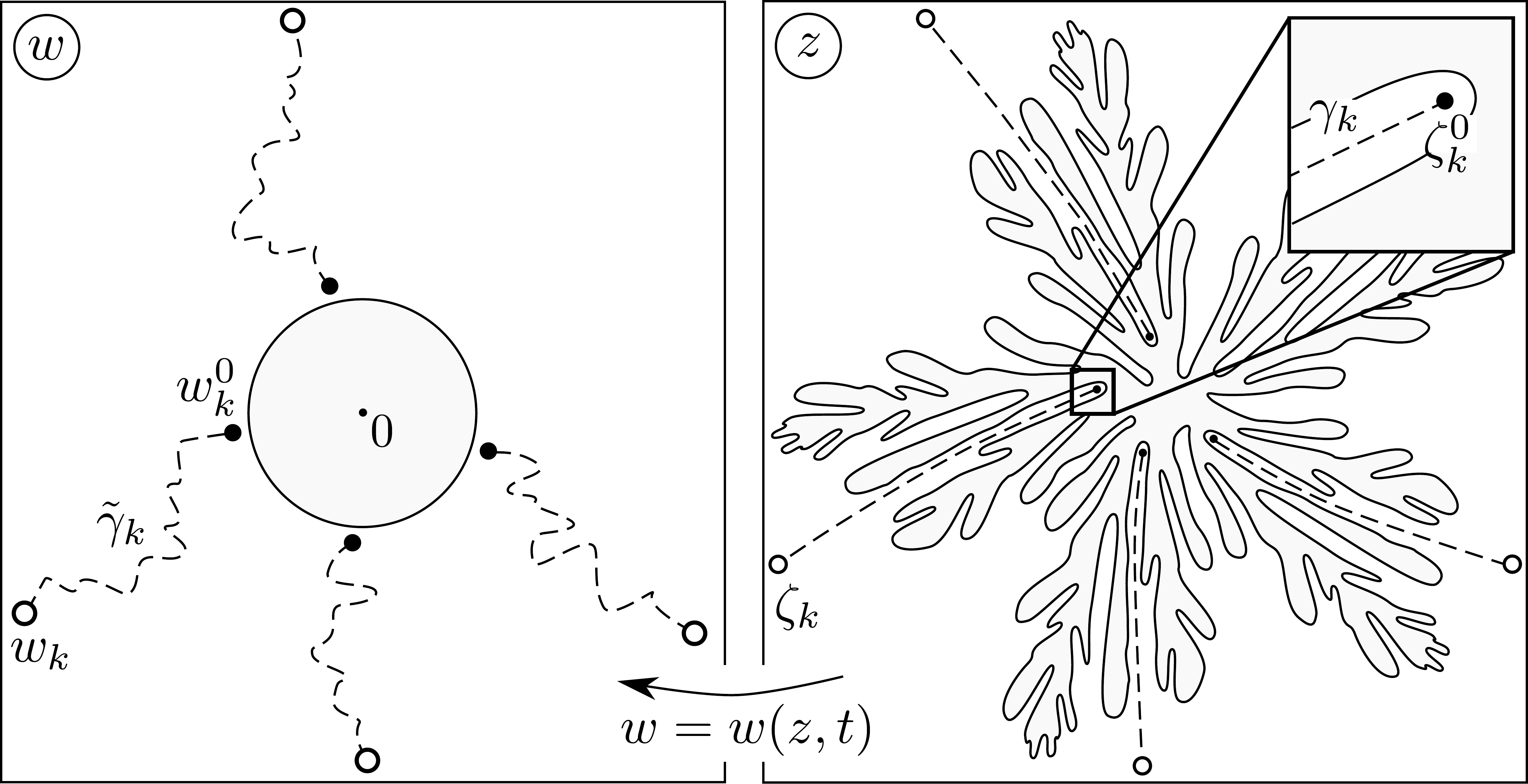}
\caption{\label{cft_domain}
The time dependent conformal map $w(z,t)$ from the exterior of the domain $D_t$ to the complement of the unit disk $\mathbb D$ in the auxiliary $w$ plane. The dashed lines in the $z$ plane represent the fjord centerlines $\gamma_k$ bounded by the endpoints $\zeta^0_k\equiv z(w_k^0(t),t)=const$ and $\zeta_k=z(w_k(t),t)$. The curve $\tilde \gamma_k$ in the $w$ plane with the endpoints $w_k$ and $w^0_k$ is the image of $\gamma_k$ under the conformal map.}
\end{figure}

\textit{Loewner-Kufarev equation.} Below, we use a Loewner-Kufarev theory to study two-dimensional growth phenomena~\cite{Loewner23,Kufarev}. Let $w(z,t):\ \mathbb C\setminus D_t\to\mathbb D$ be the time dependent conformal map from the exterior of the simply connected domain $D_t$ to the complement of the unit disk $\mathbb D$ in the auxiliary $w$ plane (see Fig.~\ref{cft_domain}). The map is unique provided the following conditions: $w(\infty,t)=\infty$, and the conformal radius $r(t)=1/w'(\infty,t)$ is a positive function of time $t$~\footnote{Here and below, prime and dot denote the partial derivatives with respect to the coordinate an time respectively.}. The growth of $D_t$ can be then represented as a Loewner chain, i.e., a sequence of conformal maps satisfying the ordinary integro-differential equation~\cite{Loewner23,Kufarev},
\begin{equation}\label{LKeq}
	\frac{dw}{dt}=w\int_0^{2\pi}\frac{d\phi}{2\pi}\frac{e^{i\phi}+w}{e^{i\phi}-w}\rho(e^{i\phi},t),\quad w(z,0)=z,
\end{equation}
where the Loewner density $\rho(w,t)$ specifies the growth process. The case when $\rho(e^{i \phi},t)=\delta(\phi-\phi_0)$ is the Dirac peak, generates the \textit{local} Loewner growth~\cite{Loewner23}. In particular, stochastic dynamics of the peak at $\phi_0$ leads to the Schramm-Loewner evolution~\cite{Schramm2000}.

The deterministic Laplacian growth is a \textit{nonlocal} process generated by the density $\rho^*(w,t)=Q/|z'(w,t)|^2$, where $z(w,t)$ is the inverse to the map $w(z,t)$, and $Q$ is the growth rate. The fluctuations around $\rho^*$, which can be written in the form~\cite{QLG2}
\begin{equation}\label{rho}
	\rho(w,t)=\frac{1}{|z'(w,t)|^2}\left[Q-2\nu\sum_{k=1}^N\Re\frac{\xi_k(t)}{w-\xi_k(t)}\right],
\end{equation}
result in the fluctuations of the normal interface velocity $v_n=|z'(e^{i\phi},t)|\rho(e^{i\phi},t)$. By $\nu$ we denoted the quanta of the growth rate, $Q$ ($Q\gg\nu$), and the points $\xi_k$ (inside the unit circle) are the collective coordinates parameterizing the noise. Due to eq.~\eqref{LKeq} these points are the poles of $z(w,t)$. In Ref.~\cite{QLG2} we argued that dissipation of the fluctuations at the interface results in the Calogero dynamics of the collective coordinates,
\begin{equation}\label{xi-mf}
	\frac{d\xi_k}{\xi_k}=-\sigma d \tau+\frac12\sum^N_{l\neq k}\frac{\xi_k+\xi_l}{\xi_k-\xi_l}d \tau,
\end{equation}
where the term $\sigma=Q/\nu+N$ produces a drift toward the origin, and $\tau(t)$ parameterizes trajectories of poles.

\textit{Stochastic Laplacian growth.} Upon quantization the Laplacian growth problem is similar to the semiclassical evolution of the electronic droplet in the quantum Hall regime~\cite{WiegmannQH,QLG1}. In the latter system the quantum effects are strong even in the semiclassical limit. To take account of quantum effects, the only mean-field trajectory~\eqref{xi-mf} should be replaced by the ensemble of trajectories, which result in the Langevin dynamics of the collective coordinates~\cite{Lacroix14}. Therefore, we consider the following  system of stochastic partial differential equations
\begin{multline}\label{dxi}
	d\log\xi_k = -\sigma dq_k-\frac{\kappa}{2}g(\xi_1,\dotsc,\bar\xi_1,\dotsc)dq_k+\\
	+\frac{1}{2}\sum_{l\neq k}^N\frac{\xi_k+\xi_l}{\xi_k-\xi_l}dq_k+\frac{1}{2}\sum_{l=1}^N\frac{\xi_k+1/\bar\xi_l}{\xi_k-1/\bar\xi_l}dq_k+i d B_k.
\end{multline}
Here, the trajectories of poles are parameterzied by the different ``times'' $q_k(t)$. The random processes, $dB_k(t)$, are determined by the mean zero and covariance $\text{Cov}[d B_k(t),d B_l(0)]=(\kappa/2)\delta_{kl}dq_k(t)$. Besides, we added the interaction between the points inside the unit circle and their ``mirror images'' outside. By $g=\xi^{-1}_k(\xi_k\p_{\xi_k}+\bar\xi_k\p_{\bar \xi_k})\log Z$ we denoted the logarithmic derivative of the correlation function $Z=\langle \prod_k\Psi(\xi_k,\bar\xi_k)\rangle_{\mathbb D}$ of the primary fields $\Psi$ with the conformal dimensions $h=-(6+\kappa)/(2\kappa)$ of the boundary Liouville conformal field theory with the central charge $c=1+3(\kappa+4)^2/(2\kappa)$.

The ansatz~\eqref{dxi} can be justified by coupling a conformal field theory (CFT) with stochastic Laplacian growth~\eqref{LKeq},~\eqref{rho} and~\eqref{dxi} in the long time asymptotic. Then, it becomes possible to express a set of \textit{martingales} in terms of conformal correlation functions~\footnote{Roughly speaking, martingales are stochastic processes whose expectation values are constant in time.}. The martingale conditions specify the drift term in eq.~\eqref{dxi}, and the distribution function of noise (cf. Ref.~\cite{BBK05}).

\textit{Correlation functions and martingales.} Let us consider a CFT outside $D_t$. It is characterized by a set of scaling operators (primary fields) $\Phi_{h}(z,\bar z)$ constituting representations of the Virasoro algebra~\cite{BPZ}. The conformal dimensions $h$ specify the transformation of the correlation functions of primary fields under conformal maps~\footnote{We only consider the spinless operators with equal holomorphic and anti-holomorphic conformal dimensions.},
\begin{equation}\label{cftcor}
	\langle \prod_k \Phi_{h_k}(z_k,\bar z_k)\rangle_{D_t}=\langle\prod_k |w'(z_k,t)|^{2h_k}\Phi_{h_k}(w_k,\bar w_k)\rangle_{\mathbb D}.
\end{equation}

The correlation functions are known to be closely related to statistical martingales of the Schramm-Loewner evolution~\cite{BBcft}. A similar approach can be used in the case of stochastic Laplacian growth, provided considerably strict constraints. Let us introduce the following normalized correlation function
\begin{equation}\label{F}
	\mathcal F_{D_t}(\{\zeta, \zeta^0\})=\frac{\langle\prod_k \Psi (\zeta_k,\bar \zeta_k)\Phi_{h}(\zeta^0_k,\bar \zeta^0_k)\rangle_{D_t}}{\langle \prod_k\Psi (\zeta_k,\bar \zeta_k)\rangle_{D_t}},
\end{equation}
where $k=1,\dotsc, N$. The positions of the fields are specified by the endpoints, $\zeta_k^0\equiv\zeta_k(0)=const$ and $\zeta_k\equiv\zeta_k(t)$, of the (dynamically generated) centerlines of the fjords of oil that separate fingers of water~\footnote{The centerlines are equidistant from the fjord edges.} (see Fig.~\ref{cft_domain}).

A stochastic evolution of the domain $D_t$ results in the Langevin dynamics of the correlation functions, which can be studied by mapping~\eqref{F} to the $w$ plane,
\begin{equation}\label{Fw}
	\mathcal F_{D_t}(\{\zeta,\zeta^0\})=\mathcal F_{\mathbb D}(\{w,w^0\})\prod_k|w'(\zeta_k^0,t)|^{2h_k},
\end{equation}
where $w_k=w(\zeta_k,t)$ and $w_k^0=w(\zeta^0_k,t)$. Let  $w\to w+\epsilon(w)$ be the infinitesimal conformal transformation generated by the Loewner chain, i.e., $\epsilon(w)=wp(w)d t$, where $p(w)$ denotes the integral over the angle in eq.~\eqref{LKeq}. Since $\Re p(e^{i\phi})<0$ this \textit{transformation does not preserve the geometry}. Taking account of $\epsilon(w)/w=\bar w\epsilon(1/\bar w)$ when $|w|=1$, and $p(1/\bar w)=-\bar p(\bar w)$, one determines the corresponding transformation of the antiholomorphic sector, namely, $\epsilon^*( w^*)=-\overline{\epsilon(w)}$~\footnote{More precisely, the CFT is defined on the Schottky double of $\mathbb C\setminus D_t$.}.

Since the conformal map changes with time stochastically, we take the It\^o derivative of~\eqref{Fw}.  To begin with, let us consider a variation of the primary field, $d\Phi_h=|w'(z,t)|^{2h}(h\epsilon'-h\overline{\epsilon'}+\epsilon\p_w-\overline{\epsilon}\p_{\bar w})\Phi_h$. Below, only the key points of the computation will be mentioned, while the technical details will be presented elsewhere.

\textit{Conformal transformations at the bottoms of fjords.} First, if the point $w$ is located in the vicinity of the unit circle, the difference $I(w)=wp'(w)-\overline{wp'(w)}$ reduces to the contour integral around $w\approx e^{i \theta}$,
\begin{equation}\label{Idef}
	I(w)=-2w\oint_{w}\frac{dv}{2\pi i}\frac{\rho(v,t)}{(v-w)^2}.
\end{equation}
Since $\rho(v,t)$ depends on the conformal map, the critical points of $z'(w,t)$ contribute to the integral. The map $z(w,t)$ is conformal outside the unit disk. Thus, all singularities of $\p z(w,t)/\p w$ are located inside the disk. Below, we consider a solution whose derivative has a form~\footnote{The domains, whose uniformization maps (more precisely, its derivative) are rational functions, are called \textit{abelian domains}~\cite{EtingofBook}.}
\begin{equation}
	z'(w,t)=r(t)\prod_{j=1}^{N'}\frac{w-u_j^c(t)}{w-u_j^s(t)},
\end{equation}
where $N'> N$ and by $u^c$ and $u^s$ we denoted the critical and singular points $z'$ correspondingly. Contrary to the deterministic Laplacian growth problem, the dynamics~\eqref{LKeq}, \eqref{rho} and~\eqref{dxi} \textit{does not preserve} the number of singular points---each time step $\delta t$ the function $z'(w,t)$ develops $N$ new poles at the points $\xi_k(t)$~\footnote{In the limit $\delta t\to 0$ the generation of poles results in the development of the branch cuts of $z'(w,t)$.}.

In the limit $r(t)\to \infty$, the mirrored pre-images, $\xi_k^0=1/\bar w^0_k$, of the fjord's tips, $\zeta^0_k=z(w_k^0,t)$, approach the unit circle, $\log(1-|\xi^0_k|)\approx-r(t)/|\alpha_k|$, where the constants $\alpha_k$ are determined by the shapes of the fjords~\cite{MW98}. Besides, the critical and singular points are separated in pairs, so that $|u^c_j-u^s_j|\sim1/r(t)$. Then, the contribution of the factor $|z'(w,t)|^{-2}$ to the integral~\eqref{Idef}, namely, $-2\Im[w_k^0z''(w^0_k)/z'(w^0_k)]\rho(w^0_k)\sim\rho(w^0_k)/r(t)$, is suppressed when compared to the contribution of the poles at $w=\xi_k$ of the density~\eqref{rho}. Thus,
\begin{equation}\label{I}
	I(w^0_k)=\frac{-2\nu}{|z'(w^0_k,t)|^2}\sum_{j=1}^N\left[\frac{w^0_k \xi_j}{(w^0_k-\xi_j)^2}-c.c.\right]+O(1/r),
\end{equation}
where $c.c.$ denotes the complex conjugated terms. The prefactor in eq.~\eqref{I} is proportional to the normal interface velocity squared, $v^*_n(e^{i \theta},t)=Q|z'(e^{i \theta},t)|^{-1}$, which becomes exponentially small ($\log|z'(w^0_k,t)|\sim r(t)/|\alpha_k|$) at the bottoms of fjords---the interface develops ``stagnation points'', that stay almost fixed during growth.

\textit{Boundary conditions.} The second key point in the computation of $d\mathcal F$ are the Neumann boundary conditions for the fields $\Phi_h(w_k,\bar w_k)=\Phi_h(e^{i \theta_k})$ located exponentially close to the unit circle,
\begin{equation}\label{bc}
	\p_n(w\bar w)^{h}\Phi_h(w,\bar w)\bigr|_{w=\exp(i \theta_k)}=0,
\end{equation}
where $\p_n=w\p_w+\bar w\p_{\bar w}$ is the normal derivative. Note, that in the coordinates, $w=\exp(i x-y)$, eq.~\eqref{bc} takes the form $\p_y \Phi_h(x,y)|_{y=0}=0$.

\textit{Transformations of fields.} The formulas~\eqref{Idef},~\eqref{I} and~\eqref{bc}, together with the equality $\Re(dw/wdt)=-\rho(w)$ for $w=\exp(i \theta)$, determine the variation of the field $\Phi^w_h(z,\bar z)\equiv |w'(z,t)|^{2h}\Phi_{h}(w,\bar w)$, under stochastic Laplacian growth,
\begin{equation}
	\frac{d\Phi^w_h(\zeta^0,\bar \zeta^0)}{|w'(\zeta^0,t)|^{2h}dt}=\left[hI(e^{i \theta})+\rho(e^{i \theta})i\p_\theta\right]\Phi_h(e^{i \theta}),
\end{equation}
where $i\p_\theta=\bar w\p_{\bar w}-w\p_w$ is the tangential derivative.

Further, let us consider a contribution of the fields $\Psi(\zeta_k,\bar \zeta_k)$ located at the endpoints $\zeta_k=z(w_k(t),t)$ of the fjord's centerlines. Since the Jacobians coming from the transformations of the fields are canceled in the numerator and denominator of the correlation function~\eqref{Fw}, the It\^o derivative of $\Psi (\xi,\bar \xi)$ reads
\begin{equation}
	d\Psi=-(\kappa/4)(\xi\p_\xi+\bar\xi\p_{\bar\xi})^2\Psi dq+\p_\xi\Psi d\xi+\p_{\bar\xi}\Psi d\bar\xi.
\end{equation}

\textit{Langevin dynamics of correlation functions.} Finally, one can neglect the difference in the normal interface velocities near stagnation points as $r(t)\to\infty$. Namely, the conformal factors $|w'(\zeta^0_k,t)|$ with $k=1,\dotsc, N$ are equal up to exponentially small corrections in $r(t)$. Then, by taking account of eq.~\eqref{dxi}, one obtains the following expression for the It\^o derivative of $\mathcal F_{D_t}(\{\zeta, \zeta^0\})$,
\begin{multline}
	\label{dF}
	d\mathcal F_{D_t}=\prod_{j=1}^N|w'(\zeta^0_j,t)|^{2h_j}\sum_{k=1}^N\left[idB_k(l_{-1}^{(k)}+\bar l_{-1}^{(k)})+\right.\\
	+\left.dq_k\left(-\frac{\kappa}{4}(l_{-1}^{(k)}+\bar l_{-1}^{(k)})^2+l^{(k,h)}_{-2}+\bar l_{-2}^{(k,-h)}\right)\right]\mathcal F_{\mathbb D},
\end{multline}
Here, we set $dq_k(t)=\nu|w'(\zeta^0_k,t)|^2dt$ (note, that all $dq_k$ with $k=1,2\dotsc$ are equal in the limit $r(t)\to\infty$), and introduced the differential operators $l_{-1}^{(k)}=\xi_k\p_{\xi_k}$, and
\begin{multline}
	l_{-2}^{(k,h)}=-\frac12\sum_{j\neq k}\frac{\xi_k+\xi_j}{\xi_k-\xi_j}\xi_j\p_{\xi_j}+\frac12\sum_j\frac{\xi_k+1/\bar\xi_j}{\xi_k-1/\bar \xi_j}\bar\xi_j\p_{\bar\xi_j}-\\
	+\frac12
\sum_j\frac{\xi_k+e^{i \theta_j}}{\xi_k-e^{i \theta_j}}i\p_{\theta_j}-\sum_{j}\frac{2h\xi_k e^{i \theta_j}}{(\xi_k-e^{i \theta_j})^2},
\end{multline}
where $h$ is the conformal dimension of $\Phi_h$. 


\textit{BPZ equations.} The primary fields of CFTs form the highest weight representations of the Virasoro algebra. These representations are not necessarily irreducible, because of the \textit{null vectors} in the Verma modules. The relevant example is the null vector at the second level, which exist whenever the weight of the highest vector, $h$, takes a value from the Kac table,
\begin{equation}
	h=h_{21}=-\frac{6+\kappa}{2\kappa},\quad \text{with}\  c=1+3\frac{(\kappa+4)^2}{2\kappa}.
\end{equation}
Note, that we consider the Virasoro algebra with $c\geq25$ (a symmetry algebra of the Liouville field theory~\cite{Nak04}), contrary to the case of the Schramm-Loewner evolution, which couples to CFTs with $c\leq1$~\cite{BBcft,BBmart}. The reason is that the differential equations for $\mathcal F$, expressing the decoupling of the null vector, matches with the expression in the right hand side of eq.~\eqref{dF} when $c\geq25$~\footnote{Note, that the poles $\xi_k$ in~\eqref{dxi} attract each other in the tangential direction, while in the case of the Schramm-Loewner evolution they are repealed.}.


The correlation functions that involve the degenerate fields (which correspond to the null vectors) satisfy linear partial differential equations called a Belavin-Polyakov-Zamolodchikov (BPZ) equations~\cite{BPZ}. In the case of the function~\eqref{F}, which involves $N$ degenerates fields $\Psi(\zeta_k,\bar\zeta_k)$, the corresponding BPZ equations read
\begin{equation}\label{BPZ}
	\left[-\frac{\kappa}{4}(l_{-1}^{(k)})^2+l^{(k,h)}_{-2}\right]\mathcal F_{\mathbb D}=0,\quad k=1,\dotsc, N,
\end{equation}
and similar equations hold for the antiholomorphic sector. Eqs.~\eqref{BPZ} allow one to recast the Langevein dynamics~\eqref{dF} of $\mathcal F$ in the form
\begin{equation}\label{dFmart}
	\frac{d\mathcal F_{D_t}(\{\zeta,\zeta^0\})}{\prod_k|w'(\zeta^0_k,t)|^{2h_k}}=\sum_{k=1}^N\left[ dB_k l_{-1}^{(k)}-dq_k\frac{\kappa}{2}\Delta^{(k)}\right]\mathcal F_{\mathbb D},
\end{equation}
where $\Delta^{(k)}=l^{(k)}_{-1}\bar l^{(k)}_{-1}$ is the Laplace operator.

\textit{Martingales of stochastic Laplacian growth.} When statistical mechanics is coupled with stochastic processes, e.g., with the Schramm-Loewner evolution, the martingales become essential ingredients for estimating probability of events, representing the conditioned correlation functions of statistical systems~\cite{BBcft,BBmart}. Roughly speaking, these are the processes whose expectation values are constant in time, i.e., the martingales satisfy stochastic differential equations without a drift term~\cite{ShreveBook}. 

The drift term in the right hand side of eq.~\eqref{dFmart} measures  the degree to which $\mathcal F_{D_t}$ fails to be a martingale (cf. Ref.~\cite{GC06}). When the drift term vanishes, i.e., the points are determined by the equation $\sum_{k=1}^N \Delta^{(k)}\mathcal F_{\mathbb D}(\{\xi,\bar\xi,e^{i \theta}\})=0$, the function~\eqref{F} represents the martingale of stochastic Laplacian growth.

\textit{Conclusion}. In this paper we proposed a relation between the martingales of stochastic Laplacian growth and correlation functions in the Liouville field theory with $c\geq25$. The correlation function~\eqref{F} determines a probability of the growth process to generate the set of fjords at the interface, whose centerlines start at $\zeta^0_k$ and pass through $\zeta_k$ as $r(t)\to\infty$. The proposed theory promises to elucidate geometrical properties of observed patterns in terms of the only parameter $\kappa$. A natural extension of this work is to study a scaling of the harmonic measure at the bottoms of fjords. It will be the first step toward the multifarctal analysis of growing clusters, which is known to be a stubborn barrier to our understanding of Laplacian growth and diffusion-limited aggregation.

Although the proposed approach provides a promising angle for understanding of nonlocal stochastic Loewner chains, much remains to be done. While the ansatz~\eqref{dxi} was justified by methods of statistical mechanics, its interpretation from the probability theory point of view is lacking. Remarkably, however, that the dynamics of the fjord's centerlines bears resemblance to the time-reversed Schramm-Loewner evolution. Another important problem is to study stochastic growth with the finite conformal radius, which serves as a ``mass'' scale. Then, the approach proposed in Refs.~\cite{BBoff09,BOOKMakarov} promises to clarify the perturbation theory for stochastic Laplacian growth.

\bibliography{biblio}{}

\begin{thebibliography}{32}%
\makeatletter
\providecommand \@ifxundefined [1]{%
 \@ifx{#1\undefined}
}%
\providecommand \@ifnum [1]{%
 \ifnum #1\expandafter \@firstoftwo
 \else \expandafter \@secondoftwo
 \fi
}%
\providecommand \@ifx [1]{%
 \ifx #1\expandafter \@firstoftwo
 \else \expandafter \@secondoftwo
 \fi
}%
\providecommand \natexlab [1]{#1}%
\providecommand \enquote  [1]{``#1''}%
\providecommand \bibnamefont  [1]{#1}%
\providecommand \bibfnamefont [1]{#1}%
\providecommand \citenamefont [1]{#1}%
\providecommand \href@noop [0]{\@secondoftwo}%
\providecommand \href [0]{\begingroup \@sanitize@url \@href}%
\providecommand \@href[1]{\@@startlink{#1}\@@href}%
\providecommand \@@href[1]{\endgroup#1\@@endlink}%
\providecommand \@sanitize@url [0]{\catcode `\\12\catcode `\$12\catcode
  `\&12\catcode `\#12\catcode `\^12\catcode `\_12\catcode `\%12\relax}%
\providecommand \@@startlink[1]{}%
\providecommand \@@endlink[0]{}%
\providecommand \url  [0]{\begingroup\@sanitize@url \@url }%
\providecommand \@url [1]{\endgroup\@href {#1}{\urlprefix }}%
\providecommand \urlprefix  [0]{URL }%
\providecommand \Eprint [0]{\href }%
\providecommand \doibase [0]{http://dx.doi.org/}%
\providecommand \selectlanguage [0]{\@gobble}%
\providecommand \bibinfo  [0]{\@secondoftwo}%
\providecommand \bibfield  [0]{\@secondoftwo}%
\providecommand \translation [1]{[#1]}%
\providecommand \BibitemOpen [0]{}%
\providecommand \bibitemStop [0]{}%
\providecommand \bibitemNoStop [0]{.\EOS\space}%
\providecommand \EOS [0]{\spacefactor3000\relax}%
\providecommand \BibitemShut  [1]{\csname bibitem#1\endcsname}%
\let\auto@bib@innerbib\@empty
\bibitem [{\citenamefont {Stanley}\ and\ \citenamefont
  {Ostrowsky}(1986)}]{StanleyBook}%
  \BibitemOpen
  \bibfield  {author} {\bibinfo {author} {\bibfnamefont {H.~E.}\ \bibnamefont
  {Stanley}}\ and\ \bibinfo {author} {\bibfnamefont {N.}~\bibnamefont
  {Ostrowsky}},\ }\href {\doibase 10.1007/978-94-009-5165-5} {\emph {\bibinfo
  {title} {On Growth and Form}}},\ Nato Science Series E:\ (\bibinfo
  {publisher} {Springer},\ \bibinfo {address} {Netherlands},\ \bibinfo {year}
  {1986})\BibitemShut {NoStop}%
\bibitem [{\citenamefont {Pelce}\ and\ \citenamefont
  {Libchaber}(1988)}]{PelceBook}%
  \BibitemOpen
  \bibfield  {author} {\bibinfo {author} {\bibfnamefont {P.}~\bibnamefont
  {Pelce}}\ and\ \bibinfo {author} {\bibfnamefont {A.}~\bibnamefont
  {Libchaber}},\ }\href@noop {} {\emph {\bibinfo {title} {Dynamics of curved
  fronts}}},\ Perspectives in Physics\ (\bibinfo  {publisher} {Academic
  Press},\ \bibinfo {address} {San Diego},\ \bibinfo {year} {1988})\BibitemShut
  {NoStop}%
\bibitem [{\citenamefont {Bensimon}\ \emph {et~al.}(1986)\citenamefont
  {Bensimon}, \citenamefont {Kadanoff}, \citenamefont {Liang}, \citenamefont
  {Shraiman},\ and\ \citenamefont {Tang}}]{BensimonRMP}%
  \BibitemOpen
  \bibfield  {author} {\bibinfo {author} {\bibfnamefont {D.}~\bibnamefont
  {Bensimon}}, \bibinfo {author} {\bibfnamefont {L.~P.}\ \bibnamefont
  {Kadanoff}}, \bibinfo {author} {\bibfnamefont {S.}~\bibnamefont {Liang}},
  \bibinfo {author} {\bibfnamefont {B.~I.}\ \bibnamefont {Shraiman}}, \ and\
  \bibinfo {author} {\bibfnamefont {C.}~\bibnamefont {Tang}},\ }\href {\doibase
  10.1103/RevModPhys.58.977} {\bibfield  {journal} {\bibinfo  {journal} {Rev.
  Mod. Phys.}\ }\textbf {\bibinfo {volume} {58}},\ \bibinfo {pages} {977}
  (\bibinfo {year} {1986})}\BibitemShut {NoStop}%
\bibitem [{\citenamefont {Witten}\ and\ \citenamefont
  {Sander}(1981)}]{WittenSanderPRL}%
  \BibitemOpen
  \bibfield  {author} {\bibinfo {author} {\bibfnamefont {T.~A.}\ \bibnamefont
  {Witten}}\ and\ \bibinfo {author} {\bibfnamefont {L.~M.}\ \bibnamefont
  {Sander}},\ }\href {\doibase 10.1103/PhysRevLett.47.1400} {\bibfield
  {journal} {\bibinfo  {journal} {Phys. Rev. Lett.}\ }\textbf {\bibinfo
  {volume} {47}},\ \bibinfo {pages} {1400} (\bibinfo {year}
  {1981})}\BibitemShut {NoStop}%
\bibitem [{\citenamefont {Richardson}(1972)}]{Richardson}%
  \BibitemOpen
  \bibfield  {author} {\bibinfo {author} {\bibfnamefont {S.}~\bibnamefont
  {Richardson}},\ }\href {\doibase 10.1017/S0022112072002551} {\bibfield
  {journal} {\bibinfo  {journal} {J. Fluid Mech.}\ }\textbf {\bibinfo {volume}
  {56}},\ \bibinfo {pages} {609} (\bibinfo {year} {1972})}\BibitemShut
  {NoStop}%
\bibitem [{\citenamefont {Mineev-Weinstein}\ \emph {et~al.}(2000)\citenamefont
  {Mineev-Weinstein}, \citenamefont {Wiegmann},\ and\ \citenamefont
  {Zabrodin}}]{WiegmannPRL}%
  \BibitemOpen
  \bibfield  {author} {\bibinfo {author} {\bibfnamefont {M.}~\bibnamefont
  {Mineev-Weinstein}}, \bibinfo {author} {\bibfnamefont {P.}~\bibnamefont
  {Wiegmann}}, \ and\ \bibinfo {author} {\bibfnamefont {A.}~\bibnamefont
  {Zabrodin}},\ }\href {\doibase 10.1103/PhysRevLett.84.5106} {\bibfield
  {journal} {\bibinfo  {journal} {Phys. Rev. Lett.}\ }\textbf {\bibinfo
  {volume} {84}},\ \bibinfo {pages} {5106} (\bibinfo {year}
  {2000})}\BibitemShut {NoStop}%
\bibitem [{\citenamefont {Alekseev}(2017{\natexlab{a}})}]{QLG1}%
  \BibitemOpen
  \bibfield  {author} {\bibinfo {author} {\bibfnamefont {O.}~\bibnamefont
  {Alekseev}},\ }\href@noop {} {\bibfield  {journal} {\bibinfo  {journal}
  {ArXiv e-prints}\ } (\bibinfo {year} {2017}{\natexlab{a}})},\ \Eprint
  {http://arxiv.org/abs/1710.08207} {arXiv:1710.08207 [cond-mat.stat-mech]}
  \BibitemShut {NoStop}%
\bibitem [{\citenamefont {Alekseev}(2017{\natexlab{b}})}]{QLG2}%
  \BibitemOpen
  \bibfield  {author} {\bibinfo {author} {\bibfnamefont {O.}~\bibnamefont
  {Alekseev}},\ }\href@noop {} {\bibfield  {journal} {\bibinfo  {journal}
  {ArXiv e-prints}\ } (\bibinfo {year} {2017}{\natexlab{b}})},\ \Eprint
  {http://arxiv.org/abs/1710.08206} {arXiv:1710.08206 [cond-mat.stat-mech]}
  \BibitemShut {NoStop}%
\bibitem [{\citenamefont {Lacroix}\ and\ \citenamefont
  {Ayik}(2014)}]{Lacroix14}%
  \BibitemOpen
  \bibfield  {author} {\bibinfo {author} {\bibfnamefont {D.}~\bibnamefont
  {Lacroix}}\ and\ \bibinfo {author} {\bibfnamefont {S.}~\bibnamefont {Ayik}},\
  }\href {\doibase 10.1140/epja/i2014-14095-8} {\bibfield  {journal} {\bibinfo
  {journal} {Eur. Phys. J. A}\ }\textbf {\bibinfo {volume} {50}},\ \bibinfo
  {pages} {95} (\bibinfo {year} {2014})}\BibitemShut {NoStop}%
\bibitem [{\citenamefont {Agam}\ \emph {et~al.}(2002)\citenamefont {Agam},
  \citenamefont {Bettelheim}, \citenamefont {Wiegmann},\ and\ \citenamefont
  {Zabrodin}}]{WiegmannQH}%
  \BibitemOpen
  \bibfield  {author} {\bibinfo {author} {\bibfnamefont {O.}~\bibnamefont
  {Agam}}, \bibinfo {author} {\bibfnamefont {E.}~\bibnamefont {Bettelheim}},
  \bibinfo {author} {\bibfnamefont {P.}~\bibnamefont {Wiegmann}}, \ and\
  \bibinfo {author} {\bibfnamefont {A.}~\bibnamefont {Zabrodin}},\ }\href
  {\doibase 10.1103/PhysRevLett.88.236801} {\bibfield  {journal} {\bibinfo
  {journal} {Phys. Rev. Lett.}\ }\textbf {\bibinfo {volume} {88}},\ \bibinfo
  {pages} {236801} (\bibinfo {year} {2002})}\BibitemShut {NoStop}%
\bibitem [{\citenamefont {L\"owner}(1923)}]{Loewner23}%
  \BibitemOpen
  \bibfield  {author} {\bibinfo {author} {\bibfnamefont {K.}~\bibnamefont
  {L\"owner}},\ }\href {\doibase 10.1007/BF01448091} {\bibfield  {journal}
  {\bibinfo  {journal} {Math. Ann.}\ }\textbf {\bibinfo {volume} {89}},\
  \bibinfo {pages} {103} (\bibinfo {year} {1923})}\BibitemShut {NoStop}%
\bibitem [{\citenamefont {Kufarev}(1941)}]{Kufarev}%
  \BibitemOpen
  \bibfield  {author} {\bibinfo {author} {\bibfnamefont {P.~P.}\ \bibnamefont
  {Kufarev}},\ }\href@noop {} {\bibfield  {journal} {\bibinfo  {journal} {Rec.
  Math. [Mat. Sbornik] N.S.}\ }\textbf {\bibinfo {volume} {13}},\ \bibinfo
  {pages} {87} (\bibinfo {year} {1941})},\ \bibinfo {note} {in
  Russian}\BibitemShut {NoStop}%
\bibitem [{Note1()}]{Note1}%
  \BibitemOpen
  \bibinfo {note} {Here and below, prime and dot denote the partial derivatives
  with respect to the coordinate an time respectively.}\BibitemShut {Stop}%
\bibitem [{\citenamefont {Schramm}(2000)}]{Schramm2000}%
  \BibitemOpen
  \bibfield  {author} {\bibinfo {author} {\bibfnamefont {O.}~\bibnamefont
  {Schramm}},\ }\href {\doibase 10.1007/BF02803524} {\bibfield  {journal}
  {\bibinfo  {journal} {Isr. J. Math.}\ }\textbf {\bibinfo {volume} {118}},\
  \bibinfo {pages} {221} (\bibinfo {year} {2000})}\BibitemShut {NoStop}%
\bibitem [{Note2()}]{Note2}%
  \BibitemOpen
  \bibinfo {note} {Roughly speaking, martingales are stochastic processes whose
  expectation values are constant in time.}\BibitemShut {Stop}%
\bibitem [{\citenamefont {Bauer}\ \emph {et~al.}(2005)\citenamefont {Bauer},
  \citenamefont {Bernard},\ and\ \citenamefont {Kyt{\"o}l{\"a}}}]{BBK05}%
  \BibitemOpen
  \bibfield  {author} {\bibinfo {author} {\bibfnamefont {M.}~\bibnamefont
  {Bauer}}, \bibinfo {author} {\bibfnamefont {D.}~\bibnamefont {Bernard}}, \
  and\ \bibinfo {author} {\bibfnamefont {K.}~\bibnamefont {Kyt{\"o}l{\"a}}},\
  }\href {\doibase 10.1007/s10955-005-7002-5} {\bibfield  {journal} {\bibinfo
  {journal} {J. Stat. Phys.}\ }\textbf {\bibinfo {volume} {120}},\ \bibinfo
  {pages} {1125} (\bibinfo {year} {2005})}\BibitemShut {NoStop}%
\bibitem [{\citenamefont {Belavin}\ \emph {et~al.}(1984)\citenamefont
  {Belavin}, \citenamefont {Polyakov},\ and\ \citenamefont
  {Zamolodchikov}}]{BPZ}%
  \BibitemOpen
  \bibfield  {author} {\bibinfo {author} {\bibfnamefont {A.~A.}\ \bibnamefont
  {Belavin}}, \bibinfo {author} {\bibfnamefont {A.~M.}\ \bibnamefont
  {Polyakov}}, \ and\ \bibinfo {author} {\bibfnamefont {A.~B.}\ \bibnamefont
  {Zamolodchikov}},\ }\href {\doibase
  http://dx.doi.org/10.1016/0550-3213(84)90052-X} {\bibfield  {journal}
  {\bibinfo  {journal} {Nucl. Phys. B}\ }\textbf {\bibinfo {volume} {241}},\
  \bibinfo {pages} {333} (\bibinfo {year} {1984})}\BibitemShut {NoStop}%
\bibitem [{Note3()}]{Note3}%
  \BibitemOpen
  \bibinfo {note} {We only consider the spinless operators with equal
  holomorphic and anti-holomorphic conformal dimensions.}\BibitemShut {Stop}%
\bibitem [{\citenamefont {Bauer}\ and\ \citenamefont
  {Bernard}(2003{\natexlab{a}})}]{BBcft}%
  \BibitemOpen
  \bibfield  {author} {\bibinfo {author} {\bibfnamefont {M.}~\bibnamefont
  {Bauer}}\ and\ \bibinfo {author} {\bibfnamefont {D.}~\bibnamefont
  {Bernard}},\ }\href {\doibase 10.1007/s00220-003-0881-x} {\bibfield
  {journal} {\bibinfo  {journal} {Comm. Math. Phys.}\ }\textbf {\bibinfo
  {volume} {239}},\ \bibinfo {pages} {493} (\bibinfo {year}
  {2003}{\natexlab{a}})}\BibitemShut {NoStop}%
\bibitem [{Note4()}]{Note4}%
  \BibitemOpen
  \bibinfo {note} {The centerlines are equidistant from the fjord
  edges.}\BibitemShut {Stop}%
\bibitem [{Note5()}]{Note5}%
  \BibitemOpen
  \bibinfo {note} {More precisely, the CFT is defined on the Schottky double of
  $\protect \mathbb C\setminus D_t$.}\BibitemShut {Stop}%
\bibitem [{Note6()}]{Note6}%
  \BibitemOpen
  \bibinfo {note} {The domains, whose uniformization maps (more precisely, its
  derivative) are rational functions, are called \protect \textit {abelian
  domains}~\cite {EtingofBook}.}\BibitemShut {Stop}%
\bibitem [{Note7()}]{Note7}%
  \BibitemOpen
  \bibinfo {note} {In the limit $\delta t\to 0$ the generation of poles results
  in the development of the branch cuts of $z'(w,t)$.}\BibitemShut {Stop}%
\bibitem [{\citenamefont {Ponce~Dawson}\ and\ \citenamefont
  {Mineev-Weinstein}(1998)}]{MW98}%
  \BibitemOpen
  \bibfield  {author} {\bibinfo {author} {\bibfnamefont {S.}~\bibnamefont
  {Ponce~Dawson}}\ and\ \bibinfo {author} {\bibfnamefont {M.}~\bibnamefont
  {Mineev-Weinstein}},\ }\href {\doibase 10.1103/PhysRevE.57.3063} {\bibfield
  {journal} {\bibinfo  {journal} {Phys. Rev. E}\ }\textbf {\bibinfo {volume}
  {57}},\ \bibinfo {pages} {3063} (\bibinfo {year} {1998})}\BibitemShut
  {NoStop}%
\bibitem [{\citenamefont {Nakayama}(2004)}]{Nak04}%
  \BibitemOpen
  \bibfield  {author} {\bibinfo {author} {\bibfnamefont {Y.}~\bibnamefont
  {Nakayama}},\ }\href {\doibase 10.1142/S0217751X04019500} {\bibfield
  {journal} {\bibinfo  {journal} {Int. J. Mod. Phy. A}\ }\textbf {\bibinfo
  {volume} {19}},\ \bibinfo {pages} {2771} (\bibinfo {year}
  {2004})}\BibitemShut {NoStop}%
\bibitem [{\citenamefont {Bauer}\ and\ \citenamefont
  {Bernard}(2003{\natexlab{b}})}]{BBmart}%
  \BibitemOpen
  \bibfield  {author} {\bibinfo {author} {\bibfnamefont {M.}~\bibnamefont
  {Bauer}}\ and\ \bibinfo {author} {\bibfnamefont {D.}~\bibnamefont
  {Bernard}},\ }\href {\doibase 10.1016/S0370-2693(03)00189-8} {\bibfield
  {journal} {\bibinfo  {journal} {Phys. Lett. B}\ }\textbf {\bibinfo {volume}
  {557}},\ \bibinfo {pages} {309} (\bibinfo {year}
  {2003}{\natexlab{b}})}\BibitemShut {NoStop}%
\bibitem [{Note8()}]{Note8}%
  \BibitemOpen
  \bibinfo {note} {Note, that the poles $\xi _k$ in~\protect \textup {\hbox
  {\mathsurround \z@ \protect \normalfont (\ignorespaces \ref {dxi}\unskip
  \@@italiccorr )}} attract each other in the tangential direction, while in
  the case of the Schramm-Loewner evolution they are repealed.}\BibitemShut
  {Stop}%
\bibitem [{\citenamefont {Karatzas}\ and\ \citenamefont
  {Shreve}(1998)}]{ShreveBook}%
  \BibitemOpen
  \bibfield  {author} {\bibinfo {author} {\bibfnamefont {I.}~\bibnamefont
  {Karatzas}}\ and\ \bibinfo {author} {\bibfnamefont {S.}~\bibnamefont
  {Shreve}},\ }\href {\doibase 10.1007/978-1-4612-0949-2} {\emph {\bibinfo
  {title} {Brownian Motion and Stochastic Calculus}}},\ \bibinfo {edition}
  {2nd}\ ed.,\ \bibinfo {series} {Graduate Texts in Mathematics}, Vol.\
  \bibinfo {volume} {113}\ (\bibinfo  {publisher} {Springer},\ \bibinfo
  {address} {New York},\ \bibinfo {year} {1998})\BibitemShut {NoStop}%
\bibitem [{\citenamefont {Gamsa}\ and\ \citenamefont {Cardy}(2006)}]{GC06}%
  \BibitemOpen
  \bibfield  {author} {\bibinfo {author} {\bibfnamefont {A.}~\bibnamefont
  {Gamsa}}\ and\ \bibinfo {author} {\bibfnamefont {J.}~\bibnamefont {Cardy}},\
  }\href {http://stacks.iop.org/0305-4470/39/i=41/a=S12} {\bibfield  {journal}
  {\bibinfo  {journal} {Journal of Physics A: Mathematical and General}\
  }\textbf {\bibinfo {volume} {39}},\ \bibinfo {pages} {12983} (\bibinfo {year}
  {2006})}\BibitemShut {NoStop}%
\bibitem [{\citenamefont {Bauer}\ \emph {et~al.}(2009)\citenamefont {Bauer},
  \citenamefont {Bernard},\ and\ \citenamefont {Cantini}}]{BBoff09}%
  \BibitemOpen
  \bibfield  {author} {\bibinfo {author} {\bibfnamefont {M.}~\bibnamefont
  {Bauer}}, \bibinfo {author} {\bibfnamefont {D.}~\bibnamefont {Bernard}}, \
  and\ \bibinfo {author} {\bibfnamefont {L.}~\bibnamefont {Cantini}},\ }\href
  {\doibase 10.1088/1742-5468/2009/07/P07037} {\bibfield  {journal} {\bibinfo
  {journal} {J. Stat. Mech.}\ }\textbf {\bibinfo {volume} {2009}},\ \bibinfo
  {pages} {P07037} (\bibinfo {year} {2009})}\BibitemShut {NoStop}%
\bibitem [{\citenamefont {Makarov}\ and\ \citenamefont
  {Smirnov}(2010)}]{BOOKMakarov}%
  \BibitemOpen
  \bibfield  {author} {\bibinfo {author} {\bibfnamefont {N.}~\bibnamefont
  {Makarov}}\ and\ \bibinfo {author} {\bibfnamefont {S.}~\bibnamefont
  {Smirnov}},\ }in\ \href@noop {} {\emph {\bibinfo {booktitle} {XVIth
  International Congress on Mathematical Physics}}},\ Vol.~\bibinfo {volume}
  {40},\ \bibinfo {editor} {edited by\ \bibinfo {editor} {\bibfnamefont
  {P.}~\bibnamefont {Exner}}}\ (\bibinfo  {publisher} {World Sci. Publ.},\
  \bibinfo {address} {Hackensack, NJ},\ \bibinfo {year} {2010})\ pp.\ \bibinfo
  {pages} {362--371}\BibitemShut {NoStop}%
\bibitem [{\citenamefont {Varchenko}\ and\ \citenamefont
  {Etingof}(1992)}]{EtingofBook}%
  \BibitemOpen
  \bibfield  {author} {\bibinfo {author} {\bibfnamefont {A.~N.}\ \bibnamefont
  {Varchenko}}\ and\ \bibinfo {author} {\bibfnamefont {P.~I.}\ \bibnamefont
  {Etingof}},\ }\href@noop {} {\emph {\bibinfo {title} {Why the Boundary of a
  Round Drop Becomes a Curve of Order Four}}},\ University Lecture Series\
  (\bibinfo  {publisher} {American Mathematical Society},\ \bibinfo {address}
  {Providence, R.I.},\ \bibinfo {year} {1992})\BibitemShut {NoStop}%
\end{thebibliography}%

\end{document}